\documentclass[twocolumn,preprintnumbers,nofootinbib,prd,superscriptaddress,groupedaddress,aps]{revtex4-1}

\usepackage{graphicx,amssymb,amsmath,amsthm,amsfonts,epstopdf,epsfig,epsf}
\usepackage[linktocpage]{hyperref}
\usepackage[usenames]{color}
\usepackage{epstopdf}

\usepackage{aas_macros}
\usepackage{bm}
\usepackage{dcolumn}
\usepackage[latin1]{inputenc}
\usepackage{latexsym}
\usepackage{rotating}
\usepackage{color}
\usepackage{longtable}
\usepackage{enumerate}
\usepackage{tensor}
\usepackage{stmaryrd}

\usepackage{url}
\def\pa{\partial}

\def\nn{\nonumber}

\def\th{\vartheta}
\def\cQ{{\cal Q}}
\newcommand{\ben}{\begin{enumerate}}
\newcommand{\een}{\end{enumerate}}

\def\be{\begin{equation}}
\def\ee{\end{equation}}
\newcommand{\beq}{\begin{eqnarray}}
\newcommand{\eeq}{\end{eqnarray}} 
\newcommand{\ba}{\begin{align}}
\newcommand{\ea}{\end{align}}

\def\nn{\nonumber}

\def\ba{\bar{a}}

\begin{document}
\preprint{CERN-TH-2017-082}

\title{Superradiance in rotating stars and pulsar-timing constraints on dark photons}

\author{
Vitor Cardoso$^{1,2}$,
Paolo Pani$^{3,1}$,
Tien-Tien Yu$^{4}$
}
\affiliation{${^1}$ CENTRA, Departamento de F\'{\i}sica, Instituto Superior T\'ecnico -- IST, Universidade de Lisboa -- UL,
Avenida Rovisco Pais 1, 1049 Lisboa, Portugal}
\affiliation{${^2}$ Perimeter Institute for Theoretical Physics, 31 Caroline Street North
Waterloo, Ontario N2L 2Y5, Canada}
\affiliation{$^{3}$ Dipartimento di Fisica, ``Sapienza'' Universit\`a di Roma \& Sezione INFN Roma1, P.A. Moro 5, 00185, Roma, Italy}
\affiliation{${^4}$ Theoretical Physics Department, CERN, CH-1211 Geneva 23, Switzerland}
\begin{abstract}
In the presence of massive bosonic degrees of freedom, rotational superradiance can trigger an instability that spins down black holes. This leads to peculiar gravitational-wave signatures and distribution in the spin-mass plane, which in turn can impose stringent constraints on ultralight fields. Here, we demonstrate that there is an analogous spindown effect for {\it conducting stars}.
We show that rotating stars amplify low frequency electromagnetic waves, and that this effect is largest when the time scale for conduction within the star is of the order of a light crossing time. This has interesting consequences for dark photons, as massive dark photons would cause stars to spin down due to superradiant instabilities. The time scale of the spindown depends on the mass of the dark photon, and on the rotation rate, compactness, and conductivity of the star. Existing measurements of the spindown rate of pulsars place \emph{direct} constraints on models of dark sectors. Our analysis suggests that dark photons of mass $m_V\sim 10^{-12}\,{\rm eV}$ are excluded by pulsar-timing observations. These constraints also exclude superradiant instabilities triggered by dark photons as an explanation for the spin limit of observed pulsars.
\end{abstract}

\maketitle

%%%%%%%%%%%%%%%%%%%%%%%%%%%%%%%%%%%%%%%%%%%%%%%
\section{Introduction}
%%%%%%%%%%%%%%%%%%%%%%%%%%%%%%%%%%%%%%%%%%%%%%%
%
The nature of dark matter is one of the biggest open questions in physics.
Broadly speaking, there are two approaches to explain the gravitational anomalies that indicate the existence of dark matter. The first is to change the way that gravity works on large scales while preserving the short-distance behavior, e.g. MOND~\cite{Milgrom:1983ca}. However, theories of modified gravity still require the addition of a dark matter particle to explain large scale structure~\cite{Dodelson:2011qv,Famaey:2011kh}. The second approach postulates that the gravitational anomalies are due to dark matter. The most popular of these explanations advocates the existence of new degrees of freedom beyond the Standard Model (SM) that form a dark sector. 
%%%%%%%%%%%%%%%%%%%%%%%%%%%%%%%%%%%%%%%%%%%%%%%%%%%%%%%%%%%%%
\subsection{Ultralight bosonic fields}
%%%%%%%%%%%%%%%%%%%%%%%%%%%%%%%%%%%%%%%%%%%%%%%%%%%%%%%%%%%%%
Some popular candidates for dark sector matter are ultralight bosonic fields. Indeed, bosonic fields are a generic feature of many theories~\cite{Arvanitaki:2009fg,Abel:2008ai}. 
A well-motivated scalar candidate is the QCD axion, a light bosonic degree of freedom introduced in physics to explain the smallness of the neutron electric dipole moment, years before the dark matter problem was fully appreciated~\cite{Peccei:1977hh,Weinberg:1977ma,Wilczek:1977pj}. In addition, a plethora of new light scalars was predicted to arise in the String Axiverse~\cite{Arvanitaki:2009fg}, making them important potential dark matter candidates. 

Vector candidates are equally well-motivated. Additional $U(1)$ gauge sectors arise in many string-motivated extensions to the SM~\cite{Abel:2008ai,Jaeckel:2010ni,Essig:2013lka}. 
In these scenarios, there can be extra degrees of freedom which are charged under both the $U(1)$ hypercharge of the SM and a ``hidden" $U(1)'$, known as dark photons.  This has motivated the study of kinetic mixing of the hidden sector with the SM. Many of these searches have been focused on eV-GeV scales using direct detection and low-energy accelerator experiments (see e.g.~\cite{Alexander:2016aln} for a summary of current efforts). 
%
%%%%%%%%%%%%%%%%%%%%%%%%%%%%%%%%%%%%%%%%%%%%%%%%%%%%%%%%%%%%%
\subsection{Superradiant instabilities and ultralight fields}
%%%%%%%%%%%%%%%%%%%%%%%%%%%%%%%%%%%%%%%%%%%%%%%%%%%%%%%%%%%%%
However, ultralight (i.e., sub-eV) fields which are weakly coupled to SM particles are difficult to probe with traditional colliding beam, fixed-target, and direct detection experiments. Instead, one can search for their imprints through their gravitational effects. 
A promising mechanism to probe bosonic fields is rotational superradiance~\cite{ZeldovichCylinder,Teukolsky:1974yv,Bekenstein:1998nt,Brito:2015oca}.
Superradiance affects all known free, bosonic fields and has been well-studied for black holes. In this context, low-frequency wavepackets of bosonic fields are amplified upon scattering off rotating black holes, when the frequency of the field wave satisfies $\omega<m\Omega$, where $m$ is the azimuthal number and $\Omega$ is the angular velocity at the event horizon~\cite{Brito:2015oca}.
When the bosonic field is massive, the effects of superradiance turn the entire system unstable~\cite{Damour:1976kh,Detweiler:1980uk,Cardoso:2005vk,Pani:2012bp,Pani:2012vp,Witek:2012tr,Endlich:2016jgc,Brito:2013wya}, and the instability gives rise to a slowly-spinning black hole surrounded by a cloud of bosonic field [cf.\ Ref.~\cite{Brito:2015oca} for an overview]. This cloud has a time-dependent quadrupole moment, and slowly dissipates through gravitational waves producing a monochromatic signal, which is a promising channel and smoking gun for new physics~\cite{Arvanitaki:2014wva,Brito:2014wla,Yoshino:2014wwa,Arvanitaki:2016fyj}. Furthermore, because superradiance drives the spin down, observations (either in the electromagnetic or gravitational-wave spectrum) of the spin-mass diagram of black holes may also bring convincing evidence for new physics~\cite{Arvanitaki:2010sy}.
Finally, it is also possible that superradiant effects are directly observable through enhanced scattering of electromagnetic or gravitational waves~\cite{Rosa:2015hoa,Rosa:2016bli}, or 
even through instabilities triggered in interstellar plasma environments surrounding black holes~\cite{Pani:2013hpa,Conlon:2017hhi}.
%
%%%%%%%%%%%%%%%%%%%%%%%%%%%%%%%%%%%%%%%%%%%%%%%%%%%%%%%%%%%%%
\subsection{Superradiance in stars}
%%%%%%%%%%%%%%%%%%%%%%%%%%%%%%%%%%%%%%%%%%%%%%%%%%%%%%%%%%%%%
%
However, superradiant effects are not limited to rotating black holes and in fact can appear in any classical system that is able to absorb radiation~\cite{ZeldovichCylinder,Bekenstein:1998nt,Richartz:2013hza,Cardoso:2015zqa,Brito:2015oca,Endlich:2016jgc}.
In this work, we show that superradiance also occurs in the presence of rotating and conducting spheres, and most notably in (rotating) stars with nonzero conductivity. 
This seemingly classical problem in electromagnetism has never -- to the best of our knowledge -- been worked out.
We find that rotating stars amplify low-frequency photons, whenever their frequency satisfies the usual superradiant condition, $\omega<m\Omega$, where now $\Omega$ is the rotational velocity of the fluid.

These superradiant effects may have interesting implications for theories of dark photons as well as more complicated hidden sector theories. We find that massive dark photons trigger an instability of rotating and conducting stars, analogous to the black hole case. 
Furthermore, the superradiant effects may be entirely contained within the dark sector, but have observable consequences that are worthy of further investigation.
The most direct signature of this scenario is the spindown of pulsars due to the superradiant instability. As we discuss, existing pulsar-timing measurements of the spindown rate of pulsars already constrain these models.
Because these pulsar-timing constraints are rather stringent, they also exclude superradiant instabilities triggered by dark photons as an alternative explanation for the spin limit of observed pulsars (cf., e.g., Refs.~\cite{Lattimer:2004pg,2005ASPC..328..317A,Patruno:2011sj,Mahmoodifar:2013quw,Haskell:2015iia} for a discussion on proposed limiting mechanisms on the spin of pulsars).
Throughout this work, we use $G=c=1$ units and unrationalized Gaussian units for the charge.
%
%%%%%%%%%%%%%%%%%%%%%%%%%%%%%%%%%%%%%%%%%%%%%%%
\section{Setup}
%%%%%%%%%%%%%%%%%%%%%%%%%%%%%%%%%%%%%%%%%%%%%%%
%%%
\subsection{Maxwell and Proca theory in curved spacetime}
%%%
To understand the effects of superradiance in stars, we work with the theory involving one vector
field $A_{\mu}$ with mass $m_V=\mu_V\hbar$,
\begin{widetext}
\be
S =\int d^4x \sqrt{-g} 
 \left( \frac{R}{16\pi} - \frac{1}{4}F^{\mu\nu}F_{\mu\nu} - \frac{\mu_V^2}{2}A_{\nu}A^{\nu}+4\pi  j^\mu A_\mu\right)+S_{\rm matter}\,,\label{action}
\ee
\end{widetext}
where $F_{\mu\nu} \equiv \nabla_{\mu}A_{\nu} - \nabla_{\nu} A_{\mu}$ is the field strength.
The vector $A_{\mu}$ can describe either Maxwell theory with the standard massless photon, in which case $\mu_V=0$, or a Proca theory in which the vector field is massive.
We will show below that in both cases there are nontrivial superradiant effects around rotating stars.
The theory above is a toy model designed to capture the main features of a general relativistic theory where a (possibly hidden) vector field is minimally coupled to the geometry.

The resulting field equations are
\begin{eqnarray}
&&\nabla_{\nu} F^{\mu\nu}+\mu_V^2A^\mu =4\pi  j^{\mu} \,, \label{proca}\\
&&G^{\mu \nu}=8\pi{T^{\mu\nu}_{\rm matter}} +16\pi\left(\frac{1}{2}F^{\mu}_{\,\,\alpha}F^{\nu\alpha}- \frac{1}{8}F^{\alpha\beta}F_{\alpha\beta}g^{\mu\nu}\right.\nonumber\\
&&\left.  - \frac{1}{4}\mu_V^2A_{\alpha}A^{\alpha}g^{\mu\nu}
+\frac{\mu_V^2}{2}A^{\mu}A^{\nu}-4\pi  j^{(\mu}A^{\nu)}\right)\,.
\end{eqnarray}
where $G_{\mu\nu}$ is the Einstein tensor and $T^{\mu\nu}_{\rm matter}$ is the standard stress-energy tensor of matter fields, the latter being collectively described by $S_m$ in action~\eqref{action}.

%%%%%%%%%%%%%%%%%%%%%%%%%%%%%%%%%%%%%%%%%%%%%%%%%%%%%%%%%%
\subsection{Background: slowly-rotating, conducting star}
%%%%%%%%%%%%%%%%%%%%%%%%%%%%%%%%%%%%%%%%%%%%%%%%%%%%%%%%%%

Because the star is assumed to be uncharged, fluctuations in the vector $A_\mu$ affect the geometry only at the quadratic order. Thus, to linear order, we can consider a standard general relativity background as a fixed geometry, around which the vector field evolves. 
We will always neglect backreaction of the vector field on the geometry. This is a reasonable approximation for all known astrophysical setups. 

We consider a slowly-spinning star and neglect quadratic- or higher-order corrections in the spin. To linear order in the spin, the background line element is described by
\be
ds^2=-F(r) dt^2+\frac{dr^2}{B(r)}-2r^2 \zeta(r) \sin^2\vartheta dt d\varphi +r^2d\Omega^2\,, \label{metric}
\ee
and the star's four velocity reads
\begin{equation}
u^\mu=F^{-1/2}\left\{1,0,0,\Omega\right\}\,, 
\end{equation}
where $\Omega$ is the rotational velocity of the fluid. The slow-rotation approximation requires $\Omega\ll \Omega_K$, with 
%%%
\begin{equation}
 \Omega_K:=\sqrt{\frac{M}{R^3}} \label{OmegaK}
\end{equation}
%%%%
being the mass-shedding frequency, whereas $M$ and $R$ are the star's mass and radius, respectively.

In the exterior, $F=B=1-2M/r$ and $\zeta=2J/r^3$, where  $J$ is the angular momentum of the star. The interior depends on the type of matter and it is described by the classical Tolman-Oppenheimer-Volkoff equations for a perfect fluid with $ T^{\mu\nu}_{\rm matter}=(P+\rho)u^\mu u^\nu+P g^{\mu\nu}$, namely
%%%
\begin{eqnarray}
 \Phi '&=& \frac{2 \left({\cal M}+4\pi r^3 P\right)}{r (r-2 {\cal M})}\,,\qquad {\cal M}'= 4\pi r^2 \rho \,,\\
 P'&=&-\frac{(P+\rho ) \left({\cal M}+4\pi r^3 P\right)}{r (r-2 {\cal M})}\,,\\
 {\varpi}''&=& \frac{4\pi  r (P+\rho) \left(r {\varpi}'+4 {\varpi}\right)}{r-2 {\cal M}}-\frac{4}{r} {\varpi}'\,,
\end{eqnarray}
%%%
where we defined $F=e^{2\Phi}$, $B=1-2{\cal M}(r)/r$ and  $\varpi:=\Omega-\zeta(r)$. Assuming a barotropic equation of state in the form $P=P(\rho)$, these equations can be integrated numerically with standard methods. For simplicity, we will focus on backgrounds describing a constant density, perfect-fluid star. In this case, the static part of the metric has an exact solution, 
\beq
{\cal M}&=&\frac{4\pi}{3}\rho r^3\,,\,\,\,P=\rho\left(\frac{\sqrt{1-2Mr^2/R^3}-\sqrt{1-2M/R}}{3\sqrt{1-2M/R}-\sqrt{1-2Mr^2/R^3}}\right)\,,\nonumber\\
e^{\Phi}&=&\frac{3}{2}\sqrt{1-2M/R}-\frac{1}{2}\sqrt{1-2Mr^2/R^3}\,,
\eeq
where $\rho=3M/(4\pi R^3)$. The equation for $\varpi$ (and therefore for $\zeta$) cannot be solved analytically for generic values of the compactness, whereas in the Newtonian limit yields $\zeta(r)=2J/R^3={\rm const}$ in the interior, which smoothly connects to $\zeta(r)=2J/r^3$ in the exterior.

Finally, the vector $A_{\mu}$ is evolving in the vicinities of an uncharged, rotating star made of material with conductivity $\sigma$ and proper charge density $\rho_{\rm EM}$. We assume that the coupling between the vector and the material is given by the constitutive Ohm's law, which in covariant form reads~\cite{Bekenstein:1998nt},
\be
j^{\alpha}=\sigma F^{\alpha\beta} u_{\beta} +\rho_{\rm EM} u^\alpha\,,
\label{eq:conductivitycurrent}
\ee
where all quantities are computed in the frame of the material whose 4-velocity is $u^{\alpha}$. This relation should be accurate for weak fields and represents the lowest order term in the family of possible couplings between the material and the vector field.
%%%%%%%%%%%%%%%%%%%%%%%%%%%%%%%%%%%%%%%%%%%%%%%%%%%%%%%%%%%%%%%%%%%%%%%%%%%%%%%%%%%%%
\vspace{-3mm}\subsection{Perturbations of a spinning, conducting star in Maxwell and Proca theory}
%%%%%%%%%%%%%%%%%%%%%%%%%%%%%%%%%%%%%%%%%%%%%%%%%%%%%%%%%%%%%%%%%%%%%%%%%%%%%%%%%%%%%
An uncharged star in electrovacuum ($A_{\mu}=0$) is a trivial solution to the previous equations. We now wish to understand linearized fluctuations around this background. We start by expanding the vector field $A_{\mu}$ in 4-dimensional vector spherical
harmonics,
\begin{widetext}
\begin{eqnarray}
A_{\mu}(t,r,\vartheta,\phi)=e^{-i\omega t}\sum_{l,m}\left( \begin{array}{cc}\left[
 \begin{array}{c} 0 \\ 0 \\
 \frac{a^{lm}(r)}{\sin\vartheta}\partial_\phi Y_{lm}\\
 -a^{lm}(r)\sin\vartheta\partial_\vartheta Y_{lm}\end{array}\right] &
 +\left[\begin{array}{c}f^{lm}(r)Y_{lm}\\
h^{lm}(r)Y_{lm} 
\\
 k^{lm}(r) \partial_\vartheta Y_{lm}\\ k^{lm}(r) \partial_\phi
 Y_{lm}\end{array}\right] \end{array}\right)\,.
\label{expansion}
\end{eqnarray}
\end{widetext}
The first term on the right-hand side has parity $(-1)^{l+1}$
and the second term has parity $(-1)^{l}$, $m$ is an azimuthal number
and $l$ is the angular number. 
Likewise, we expand the charge density in scalar spherical harmonics, $\rho_{\rm EM}(t,r,\vartheta,\phi)=e^{-i\omega t} \hat\rho_{\rm EM}(r) Y_{lm}$.

Because the background is not spherically symmetric, the above decomposition introduces couplings between polar and axial modes and between perturbations with different harmonic indices~\cite{Pani:2013pma}. 
To linear order in the spin, the coupling between polar and axial modes can be consistently neglected and one is left with an ``axial-led'' and a ``polar-led'' system of ordinary differential equations~\cite{Pani:2012vp,Pani:2012bp,Pani:2015nua}. The decoupling procedure is given in Appendix~\ref{app:slowrot}. Here, we report only the final result for the axial-led system to linear order in the spin,
\beq
&&\frac{d^2 a}{dr_*^2}+\left(\omega^2-2m\omega\zeta(r)-V\right)a=0\,, \label{axial_final}\\
&&V=F\left(\frac{l(l+1)}{r^2}+\mu_V^2-\frac{4i\pi\sigma (\omega-m\Omega)}{\sqrt{F}}\right)\,.
\eeq
where $dr/dr_*=\sqrt{BF}$. Note that, within our slow-rotation approximation, $\sigma$ can be a generic radial function. For simplicity, we take $\sigma={\rm const}$.

The polar sector is more involved and we leave it for future work. Here, we briefly mention that in the massless case ($\mu_V=0$) the polar sector can be reduced to a single second-order differential equation by using some gauge freedom, whereas in the Proca case the polar sector describes the propagation of two physical degrees of freedom, and one is left with a system of two, coupled, second-order, differential equations. In both cases, the charge density $\hat\rho_{\rm EM}(r)$ is fixed in terms of $\sigma$ and of the perturbations of the electromagnetic field by the field equations, similarly to the fluid density $\rho$ which is fixed by the Tolman-Oppenheimer-Volkoff equations in terms of the pressure once an equation of state is given. This can be also understood by the fact that an applied electric field will modify the charge distribution, even when the object is globally neutral.
% %%%%
%%%%%%%%%%%%%%%%%%%%%%%%%%%%%%%%%%%%%%%%%%%%%%%%%%%%%%%%%%%%
\vspace{-4mm}\section{Superradiant scattering from spinning and conducting stars}
%%%%%%%%%%%%%%%%%%%%%%%%%%%%%%%%%%%%%%%%%%%%%%%%%%%%%%%%%%%%
We now consider a scattering experiment. We focus on the axial sector, but the computation for the polar sector, although more technically involved, follows similarly. In the axial sector, the solutions
to Eq.~\eqref{axial_final} behave asymptotically as
\beq
a(\omega, r)&\sim & r^{l+1}\,,\qquad \hspace{2.65cm} r\to 0\,,\\
a(\omega, r)&\sim & A_{\rm in} e^{-i\omega r}+A_{\rm out} e^{+i\omega r}\,,\qquad r\to\infty\,,\label{asymptotic}
\eeq
We have selected the regular solution at the center of the star. From our conventions for the time-dependence of the fields,
it follows that this state is composed of a piece, $A_{\rm in} e^{-i\omega r}$, which is an ingoing wave and is scattered by the star, giving rise to an outgoing component $A_{\rm out} e^{+i\omega r}$. It is also easy to verify that the incoming and outgoing fluxes at infinity are proportional to $|A_{\rm in}|^2$ and $|A_{\rm out}|^2$, respectively~\cite{Teukolsky:1974yv}.
We thus define the superradiant factor 
\be
Z:=\frac{|A_{\rm out}|^2}{|A_{\rm in}|^2}-1\,.\label{amplfactor}
\ee
\begin{figure}[th]
  \includegraphics[width=0.48\textwidth]{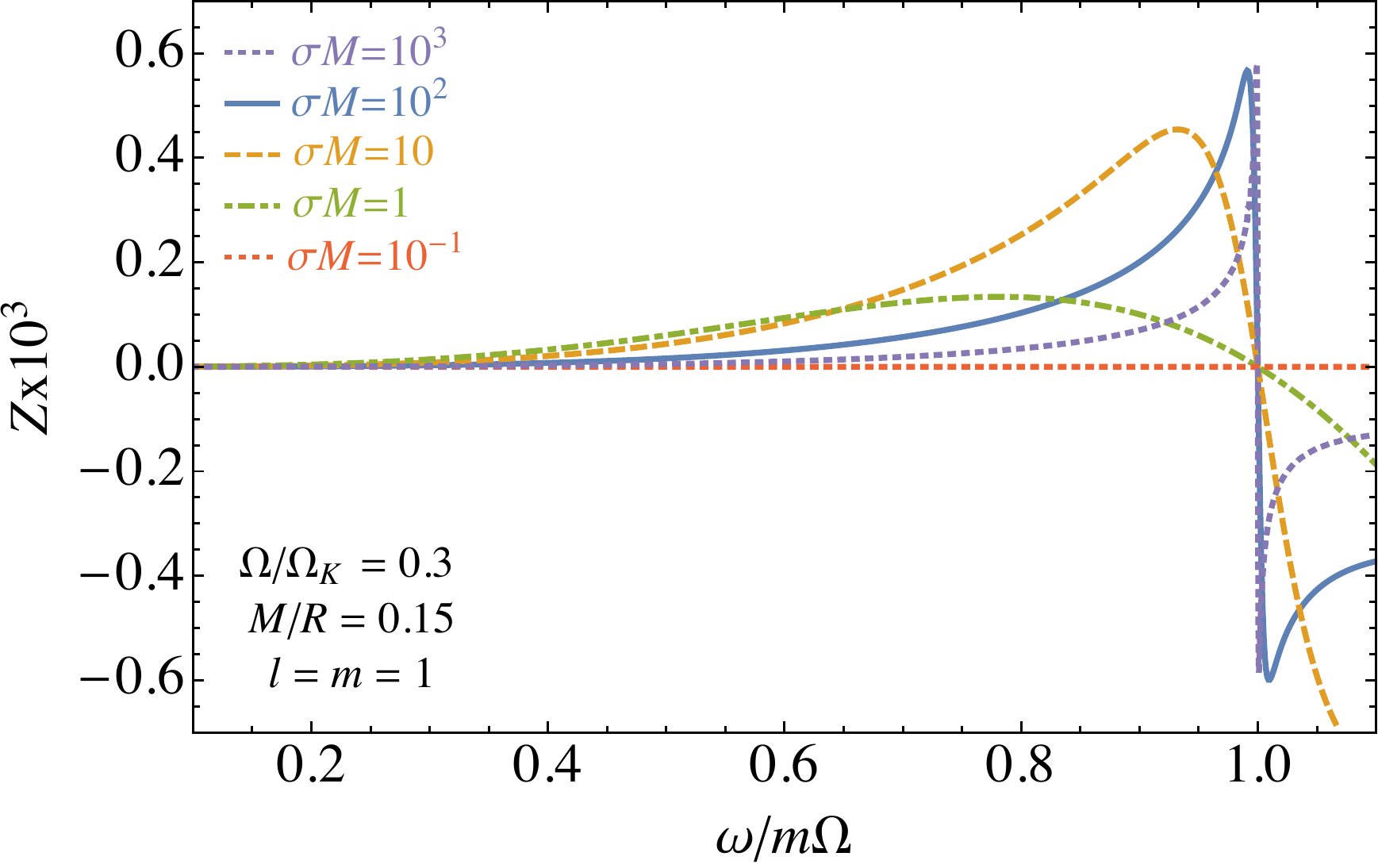}\\
\caption{Amplification factor Eq.~\eqref{amplfactor} for dipole modes $l=m=1$ as a function of the frequency, for a moderately-spinning star ($\Omega=0.3\,\Omega_K$) with compactness $M/R=0.15$ and for different values of the conductivity.} \label{fig:scattering}
\end{figure}

We have computed the superradiant factor $Z$ numerically, by integrating Eq.~\eqref{axial_final} from the center of the star, outwards to some finite but large value of the radial coordinate $r$, where the numerical solution is matched against a higher-order version of expansion \eqref{asymptotic}. The numerical results are shown in Fig.~\ref{fig:scattering}. As expected, $Z>0$ when the superradiant condition is satisfied, $\omega<m\Omega$.  The amplification factor grows with $\sigma$, until it saturates in the large-$\sigma$ limit displaying a sharp maximum at $\omega\lesssim m\Omega$. Although not shown, the amplification grows with the compactness and with the spin of the object.

We can also gain some analytical insight on the superradiant amplification. In the Newtonian limit, the external solutions are linear combinations of Bessel functions
$\sqrt{r}J_{l+1/2}(\omega r)$ and $\sqrt{r}Y_{l+1/2}(\omega r)$. In the interior, and for small conductivities, the only regular solution admissible is $\sqrt{r}J_{l+1/2}(-ir\sqrt{4im\pi\sigma\Omega-4i\pi \sigma\omega-\omega^2})$. Matching the functions and their derivatives at the surface of the star and expanding for small frequencies, we find
\be
Z=-\frac{2^{1-2l}\pi^2}{\Gamma[l+3/2]\Gamma[l+5/2]}\sigma R^2\left(\omega -m\Omega\right)(\omega R)^{2l+1}\,.
\ee
The above expression agrees remarkably well with the exact numerical result up to $M/R\sim 0.2$ and for $\sigma M\ll1$.
This relation is also interesting, as it extends an observation made in Ref.~\cite{Cardoso:2015zqa}:
one can try to naively compute the superradiant amplification factors of Kerr black holes by letting
$R=2M$ and $1/\sigma = M$, as this is now the
only possible time scale in the problem. 
With this substitution, the above relation predicts that slowly rotating black holes
in general relativity amplify $l=1$ scalar fields with
$Z=\frac{64\pi}{45}M\left(\Omega-\omega\right)(2M\omega)^3$. 
On the other hand, a matched-asymptotic expansion calculation
in full general relativity yields the same result to within an order of magnitude
(the coefficient turns out to be $8/9$ instead of $64\pi/45$)~\cite{Brito:2015oca}.
As we show in Appendix \ref{app:membrane}, one can improve on this relation by using the membrane paradigm for describing horizons~\cite{Thorne:1986iy}.
In this framework, horizons are endowed with a surface conductivity of $1/4\pi$, and a simple
Newtonian analogue recovers {\it exactly} the general relativistic prediction.

For large conductivities, we have been unable to find concise analytical expressions, but in the Newtonian limit our results are well approximated by
\be
Z=k_l\frac{\left(\omega R\right)^{2l+1}}{\sqrt{\sigma R^2\left(m\Omega-\omega\right)+c_l}}\left[1+\frac{1}{2\sigma R^2\left(m\Omega -\omega\right)}\right]^{-1}\,,\label{large_sigma_Z}
\ee
in the superradiant regime, with $k_1\sim 0.78,\,k_2\sim0.09$ and $c_1\sim2, c_2\sim 25$. 
The amplification factor is peaked at $\omega-m\Omega \sim 1/(\sigma R)$, and bounded. The analytical expression above is not very accurate close to the peak of the amplification factor, but we find numerically that, for $\sigma M\gg1$, the $l=m=1$ peak is described by
\be
Z_{\rm max} \sim (0.48-0.78 M/R) (\Omega R)^3\,,
\ee
where, interestingly, the prefactor decreases at large compactness.

%%%%%%%%%%%%%%%%%%%%%%%%%%%%%%%%%%%%%%%%%%%%%%%%%%%%%%%%%%%%
\section{Superradiant instabilities of spinning and conducting stars}
%%%%%%%%%%%%%%%%%%%%%%%%%%%%%%%%%%%%%%%%%%%%%%%%%%%%%%%%%%%%

In analogy with the black hole case, we expect that the mass term for the Proca field can lead to superradiant instabilities in conducting stars. We show this explicitly by solving the perturbation equations numerically as an eigenvalue problem, and computing the quasinormal modes of the system, $\omega_{lmn}=\omega_R+i\omega_I$, where $n$ is the overtone number. In our notation, an instability corresponds to $\omega_I>0$, and $\tau\equiv 1/\omega_I$ is the instability time scale.
The parameter space of the spectrum is large and complicated, since --~even for fixed ``quantum'' numbers $(l,m,n)$~-- it still depends on four dimensionless parameters, namely ($\mu_V M, M/R, \Omega/\Omega_K$, $\sigma M$).

\begin{figure}[th]
  \includegraphics[width=0.48\textwidth]{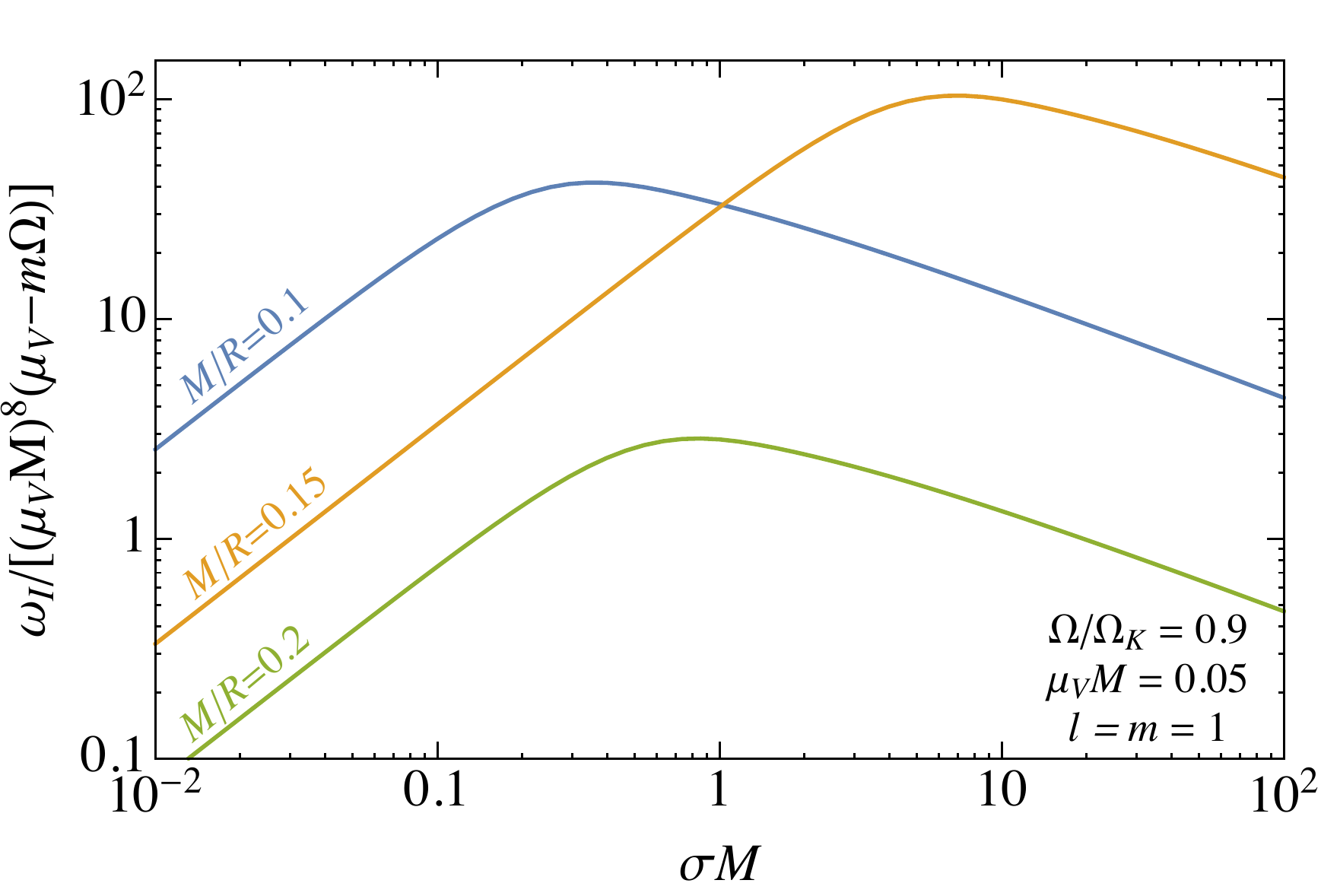}\\
\caption{The prefactor in square brackets of Eq.~\eqref{wIaxial} as a function of $\sigma M$ and for different values of the compactness at fixed $\Omega/\Omega_K=0.9$. A fit of the numerical data is consistent with Eq.~\eqref{wIaxial} with $(\alpha_1,\alpha_2)\sim(39,0.13)$, $(429,11)$ and $(4.2,0.48)$ for $M/R=0.1,0.15,0.2$, respectively.} \label{fig:instability}
\end{figure}

In the axial case, our results for the $l=m=1$ fundamental unstable mode are well approximated in the small $\mu_VM$ limit and to linear order in $\Omega/\Omega_K$ by
\begin{eqnarray}
\omega_R^2&\sim&\mu_V^2\left(1-\frac{\mu_V^2M^2}{8}\right)\,,\\
\omega_I&\sim&-\left[\alpha_1\frac{\sigma M}{\alpha_2+(\sigma M)^{3/2}}\right](\mu_VM)^8 (\mu_V-m\Omega)\,, \label{wIaxial}
\end{eqnarray}
where $\alpha_i$ are dimensionless constants that depend on the compactness and also on $\Omega$ since the combination $ \Omega/\sigma$ is not necessarily small. Besides the prefactor in square brackets in Eq.~\eqref{wIaxial}, the functional form of the superradiantly unstable modes is the same as that found for a black hole~\cite{Pani:2012bp,Pani:2012vp,Rosa:2011my,Witek:2012tr}.
The dependence of the prefactor in Eq.~\eqref{wIaxial} on $\sigma$ and $M/R$ are presented in Fig.~\ref{fig:instability}, which confirms the linear behavior in $\sigma$ at small conductivities and the $\sim\sigma^{-1/2}$ behavior at large conductivities. Furthermore, the dependence on the compactness is monotonic at small conductivities, but it is more complicated at large conductivities, in line with our findings for the amplification
factor of massless fields [see discussion around Eq.~\eqref{large_sigma_Z}].
Note that, because $\omega_R\sim \mu_V$, the small-rotation approximation together with the superradiant condition requires $\mu_V\lesssim m\Omega\ll m\Omega_K$, which implies $\mu_V M\ll 1$. To avoid the factor $(\mu_VM)^8$ in Eq.~\eqref{wIaxial} to be exceedingly small, we consider in Fig.~\ref{fig:instability} a large rotation rate, $\Omega\sim 0.9\Omega_K$, although we stress that our results are also valid for smaller values of $\Omega$. 

In Appendix~\ref{app:membrane}, we discuss a simple model that shares many features with our numerical results.

%

%%%%%%%%%%%%%%%%%%%%%%%%%%%%%%%%%%%%%%%%%%%%%%%%%%%%%%%%%%%%
\section{Phenomenological implications}
%%%%%%%%%%%%%%%%%%%%%%%%%%%%%%%%%%%%%%%%%%%%%%%%%%%%%%%%%%%%
We now discuss some potential phenomenological implications of the superradiant instability of stars. We begin with a discussion of the standard (i.e., electromagnetic) conductivity of a neutron star and then generalize the discussion to the conductivity of a hidden sector. Finally, we discuss the implications of the superradiant instability of pulsars for models of dark photons. 

%%%
\subsection{Conductivity in Maxwell theory}
%%%

The conductivity of a material can be estimated by a simple Drude model, 
%%%
\begin{equation}
 \sigma = \frac{n_e e^2 \tau_{\rm ft}}{m_e}\,, \label{sigma}
\end{equation}
where $n_e$, $e$, and $m_e$ are the number density, the charge, and the mass of the charge carriers, and $\tau_{\rm ft}$ is the mean free time between ionic collisions. The standard charge carriers are electrons and the ionic collisions are between the electrons and protons through electromagnetic interactions. The interaction between electrons and neutrons is small as it proceeds solely through the neutron magnetic moment. 

More generally, the expression for $\tau_{\rm ft}$ will depend on all possible interactions of the electron with protons within the conducting material,
\beq
\frac{1}{\tau_{\rm ft}}=\frac{\hbar^2 k_F^2}{48\pi}\left(\frac{T}{T_p}\right)^2\int_0^{2k_F} {\rm d}q \,q^2|{\cal M}|^2,
\eeq
where $k_F=\sqrt{2m_e E_F}/\hbar=(3\pi^2 n_e)^{1/3}$ is the Fermi wavenumber, $T_p=\hbar^2 k_F^2/(2m_p k_B)$ 
  is the proton Fermi temperature, $m_p$ is the proton mass, $\hbar q$ is the momentum transfer of the collision, and $k_B$ is the Boltzmann constant. $|{\cal M}|^2$ is the proton-electron scattering matrix element, and in the limit where the electron energies are much smaller than the proton mass, it is given by the Mott formula
\beq
|{\cal M}|^2=\left[\frac{4\pi e^2}{\hbar^2(q^2+k_{\rm FT}^2)}\right]^2\left(1-\frac{q^2}{4k_F^2}\right), \label{Mott}
\eeq
where $k_{\rm FT}$ is the Fermi-Thomas screening wavenumber for the system. In a neutron star, the protons are much more polarizable than electrons and so $k_{\rm FT}$ corresponds to the contribution of protons alone, i.e. 
\begin{equation}
 k_{\rm FT}^2=\frac{4 k_F m_p e^2}{\pi\hbar^2} =\frac{4m_p e^2}{\pi\hbar^2} (3\pi^2 n_e)^{1/3} \,. \label{kFT}
\end{equation}
We assume the star to be electrically neutral, $n_e=n_p$.
To first order in $k_{\rm FT}/k_F\ll1$, 
 \beq
 \frac{1}{\tau_{\rm ft}}\sim e^4\frac{\pi^2}{12\hbar^2}\left(\frac{T}{T_p}\right)^2 \frac{k_F^2}{k_{\rm FT}}.
 \label{eq:tauinv1order}
 \eeq
%%%
Together with Eq.~\eqref{sigma}, this yields~\cite{1969Natur.224..674B}
%%%
\begin{equation}
 \sigma_{\rm EM}\sim  2 \left(\frac{3}{\pi}\right)^{3/2}\frac{ \hbar^4 (m_p n_e)^{3/2}}{ e  m_p^{3} k_B^2T^2}\,, \label{final}
\end{equation}
%%%
where we included the label ``EM'' to distinguish the above electromagnetic conductivity from the hidden conductivity discussed below.

For a typical neutron star with mass density $m_p n_e\simeq 10^{13}\,{\rm g/cm}^3$ and $T\simeq 10^8\,{\rm K}$, the above formula yields $\sigma_{\rm EM}\simeq 6\times 10^{22}\,{\rm s}^{-1}$, which in our units translates to $\sigma_{\rm EM} M\simeq 10^{17}$ for a typical neutron star mass. 
In this scenario, where $\sigma_{\rm EM} M\gg 1$, we obtain from Eq.~\eqref{wIaxial} a typical instability time scale
%%%
\begin{equation}
 \tau \sim \frac{\sqrt{\sigma_{\rm EM} M}}{\alpha_1 (\mu_V M)^8 (\mu_V-\Omega)}M\simeq 10^6\,{\rm yr}\,, \quad \sigma_{\rm EM}M\gg1 \label{tauMaxwell}
\end{equation}
%%%
where in the last estimate we considered $M= 1.4 M_\odot$, $M/R=0.15$, $\alpha_1=429$, $\Omega=0.9\Omega_K$, $\mu_V M=0.05$, and $\sigma_{\rm EM} = 6\times 10^{22}\,{\rm s}^{-1}$. Therefore, even when $\sigma_{\rm EM} M\sim 10^{17}$, the instability timescale can be smaller than a typical accretion time scale, $\tau_{\rm Salpeter}\simeq 4.5\times 10^7\,{\rm yr}$. Note  that the above estimate was in the regime where the stellar angular velocity is close to the mass-shedding limit, $\Omega= 0.9\Omega_K$, and the compactness corresponds to the strongest instability [cf.\ Fig.~\ref{fig:instability}]. In this case, the superradiant instability timescale of neutron stars is actually shorter than that of nearly-extremal BHs~\cite{Dolan:2007mj}. As discussed below, the measured spin of neutron stars is at least a factor of two smaller than the mass shedding limit. Since the timescale will increase for lower angular velocities and for other values of the compactness, Eq.~\eqref{tauMaxwell} can be taken as a lower limit.

%%%%%%%%%%%%%%%%%%%%%%%%%%%%%%%%%%%%%%%%%%%%%%
\subsection{Conductivity in Hidden Sectors}
%%%%%%%%%%%%%%%%%%%%%%%%%%%%%%%%%%%%%%%%%%%%%%

We now extend the above discussion to include models of a secluded $U(1)'$~\cite{Holdom:1985ag,Pospelov:2008zw,Jaeckel:2010ni} with a massive vector boson $X$. 
For this scenario, we will consider the low-energy effective Lagrangian,
\begin{eqnarray}
{\cal{L}}_{eff}&\supset& -\frac{1}{4}F_{\mu\nu}F^{\mu\nu}-\frac{1}{4}X_{\mu\nu}X^{\mu\nu}\\
&+&\frac{\epsilon}{2}F_{\mu\nu}X^{\mu\nu}+\frac{m_X^2}{2}X_{\mu}X^{\mu}+j_{\mu}A^{\mu}\nonumber,
\end{eqnarray}
where $F_{\mu\nu}$ is the field strength of the Maxwell vector $A_\mu$, $X_{\mu\nu}$ is the field strength of the new $U(1)'$ gauge boson $X_\mu$, $m_X$ is the mass of $X$, and $\epsilon$ is the kinetic mixing between the two sectors. One can rotate away the kinetic mixing term by working in the mass basis with $A_\mu\to A_\mu+\epsilon X_\mu$, but this induces a new term $\epsilon j_{\mu}X^{\mu}$ in the Lagrangian. The physical consequence is that particles with electric charge also carry a hidden charge $\epsilon e$. Therefore, Eq.~\eqref{eq:conductivitycurrent} is modified with $\sigma F^{\mu\nu}u_{\nu}\to \sigma F^{\mu\nu}u_{\nu}+\sigma_\epsilon X^{\mu\nu}u_{\nu}$, where $\sigma_\epsilon=\epsilon\sigma_{\rm{EM}} $ to leading order in $\epsilon$.
For sub-eV $m_X$, the primary constraints on $\epsilon$ are from stellar production of the vector~\cite{An:2013yfc,An:2013yua}, precision tests of electromagnetism~\cite{Bartlett:1988yy,Betz:2013dza,Graham:2014sha}, and distortion of the cosmic microwave background (CMB) due to conversion of $\gamma\to X$~\cite{Mirizzi:2009iz}, which sets $\epsilon<{\mathcal O}(10^{-7}-10^{-5})$,  depending on $m_X$.  
One can further limit $\epsilon<{\mathcal O}(10^{-12}-10^{-8})$ by constraining the cosmic abundance of $X$ through CMB distortions due to the conversion of $X\to\gamma$~\cite{Arias:2012az,Graham:2015rva}, while proposed electromagnetic resonator technologies can potentially probe even smaller values of $\epsilon$~\cite{Chaudhuri:2014dla}. 
Thus, the effective conductivity in these models can be much smaller than in Maxwell theory, $\sigma_\epsilon\ll\sigma_{\rm EM}$.

In this context it is also relevant to estimate plasma effects, since neutron stars will be surrounded by plasma in various forms. In Maxwell theory, ordinary photons propagating in a plasma acquire an effective mass given by~\cite{Sitenko:1967}
\begin{equation}
\omega_p\hbar =\sqrt{\frac{4\pi e^2 n_{\rm plasma}}{m_e}}\approx 3\times 10^{-11}\sqrt{\frac{n_{\rm plasma}}{1\,{\rm cm^{-3}}}}\,{\rm eV}\,,\label{plasma_freq}
\end{equation}
where $n_{\rm plasma}$ is the electron number density in the plasma. In the millicharged cases, we should replace $e\to \epsilon e$ in the above equation. In the context of superradiance~\cite{Pani:2013hpa,Conlon:2017hhi}, plasma effects can be neglected as long as $\omega_p\ll \mu_V$. As discussed below, the relevant range of dark-photon masses is $\mu_V\hbar \sim 10^{-12}\,{\rm eV}$. Therefore, if $\epsilon<{\cal O}(10^{-12})$ plasma effects are negligible whenever $n_{\rm plasma}<10^{21}\,{\rm cm^{-3}}$.

We can also consider a case of a more complicated hidden sector in which the conductivity $\sigma'$ is set by the interactions between particles of opposite $U(1)'$ charge, which we denote as hidden electrons and hidden protons, with the hidden electrons serving as the charge carriers (cf., e.g., Refs.~\cite{Mohapatra:2001sx,Ackerman:mha}). Here, $j^\mu\to j'^{\mu}=\sigma' X^{\mu\nu}u_\nu+\rho' u^\mu$, which is entirely contained within the hidden sector. The calculation of $\sigma'$ requires the replacement\footnote{In the context of superradiant mechanisms, the relevant Compton wavelength of dark photons is much larger than the mean free path of the hidden electrons in the stars. Thus, the mass of the mediator has a negligible effect on the conductivity calculation.} in Eq.~\eqref{final} of $e$ by the hidden electric charge $e'$, $m_e$ and $m_n$ by the mass of hidden electrons $m_{e'}$ and nucleons $m_{n'}$, and $n_e$ by the number density of hidden electrons $n_{e'}$. 
This manifests itself in Eq.~\eqref{final} by the replacements $e\to e'$, $m_n\to m_{n'}$ and $n_{e}\to n_{e'}$, giving 
\beq
\frac{\sigma'}{\sigma_{\rm {EM}}}=\left(\frac{n_{e'} m_p}{n_e m_{p'}}\right)^{3/2}\frac{e}{e'}.
\eeq
Taking for instance $e' = 0.01e$~\cite{Ackerman:mha,Cardoso:2016olt}, $m_{p'}= 100\,{\rm TeV}$, and assuming that the mass density of hidden protons inside the star is $1\%$ ($0.1\%$) of the mass density of ordinary protons, we estimate a conductivity for hidden electrons $\sigma'\simeq10^{-16}\sigma_{\rm{EM}}~( 3\times10^{-18}\sigma_{\rm{EM}})$,  {\emph {i.e.}} $\sigma' M \simeq 10$ ($0.3$). 
In other words, models of hidden $U(1)'$ sectors above the TeV scale can have dramatically smaller values of neutron-star conductivity for the hidden electron than that of ordinary electrons, and values $\sigma' M\sim {\cal O}(1)$ are allowed. Thus, in our estimates we will consider $\sigma$ as a free parameter.

%%%%%%%%%%%%%%%%%%%%%%%%%%%%%%%%%%%
\subsection{Instability Time Scale}
%%%%%%%%%%%%%%%%%%%%%%%%%%%%%%%%%%%

\begin{figure}[th]
  \includegraphics[width=0.48\textwidth]{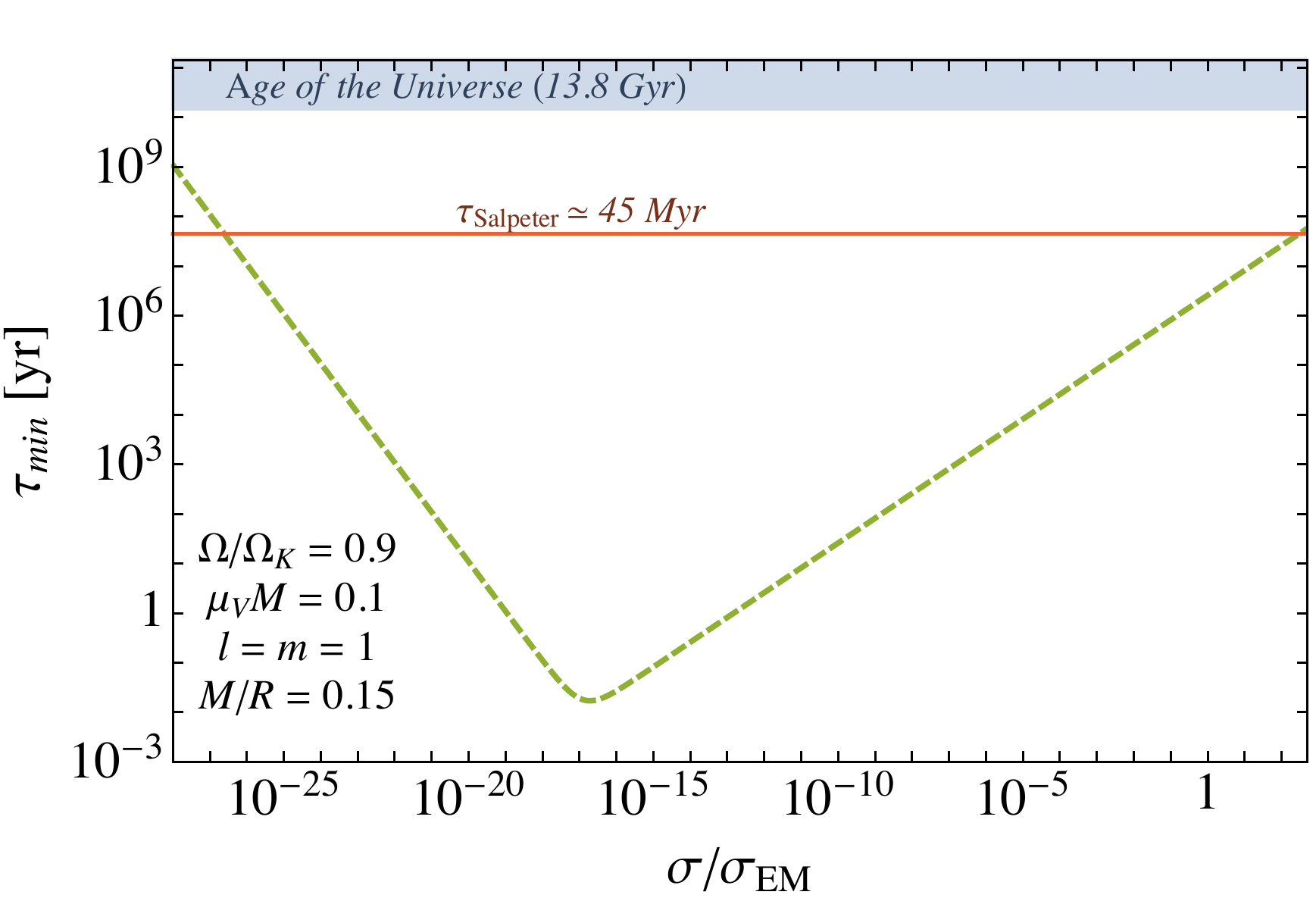}\\
\caption{Minimum time scale for the superradiant instability of a neutron star against hidden vectors in the axial sector as a function of the the ratio $\sigma/\sigma_{\rm EM}$ where $\sigma_{\rm EM}$ is the typical conductivity of ordinary electrons in a neutron star. We considered a typical neutron star with $M= 1.4 M_\odot$, $M/R=0.15$, and rotating near the mass-shedding limit, $\Omega= 0.9\Omega_K$. The minimum instability time scale corresponds to a dark photon with mass $m_V\simeq 4\times 10^{-12}\,{\rm eV}$. } \label{fig:tmin}
\end{figure}

As discussed in Refs.~\cite{Pani:2012vp,Pani:2012bp}, the minimum instability time scale $\tau\equiv 1/\omega_I$ can be estimated by computing the value of $\mu_V$ which corresponds to the maximum value of $\omega_I$. From Eq.~\eqref{wIaxial}, $d\omega_I/d\mu=0$ yields $\mu_V^{\rm min}=8\Omega/9$, which corresponds to
%%%
\begin{equation}
 \tau_{\rm min} =  \frac{387420489}{16777216 }\left(\frac{\alpha_2+(\sigma M)^{3/2}}{\alpha_1  \sigma  (mM\Omega)^9}\right)\,.
\end{equation}
%%%
The minimum instability time scale is shown in Fig.~\ref{fig:tmin} as a function of the ratio $\sigma/\sigma_{\rm EM}$ where $\sigma_{\rm EM}$ is the typical conductivity of ordinary electrons in a neutron star. As expected, $\tau_{\rm min}$ diverges both when $\sigma\to0$ and when $\sigma\to\infty$, and it displays a minimum at $\sigma\simeq 10^{-17}\sigma_{\rm EM}$, which corresponds to $\sigma M\simeq 1$. Note also that $\tau_{\rm min}$ depends strongly on $\Omega$. In Fig.~\ref{fig:tmin}, we considered the extreme case $\Omega=0.9\Omega_K$, but $\tau_{\rm min}$ roughly scales with $(\Omega_K/\Omega)^9$. Thus, the time scale for $\Omega=0.3\Omega_K$ will be roughly $5\times10^4$ times longer than that shown in Fig.~\ref{fig:tmin}.

%

%%%%%%%%%%%%%%%%%%%%%%%%%%%%%%%%%%%%%%%%%%%%%%%%%%%%%%
\subsection{Pulsar-Timing Constraints on Dark Photons}
%%%%%%%%%%%%%%%%%%%%%%%%%%%%%%%%%%%%%%%%%%%%%%%%%%%%%%
%
Various arguments~\cite{Brito:2015oca} suggest that the superradiant instability extracts angular momentum from the central object, spinning it down until the superradiant condition is saturated, $\mu_V\sim\omega_R=m\Omega$ (this was recently confirmed by the first numerical simulations\footnote{A related result was shown to hold for charged scalar perturbations of Reissner-Nordstr\"om black holes (which also exhibit superradiance~\cite{Brito:2015oca}) both perturbatively~\cite{Brito:2015oca} and in full nonlinear simulations~\cite{Bosch:2016vcp,Sanchis-Gual:2015lje}.} of massive vector fields around a spinning black hole~\cite{East:2017ovw}). 
The superradiant instability develops by extracting energy away from the spinning object and depositing it on
a bosonic condensate (or a ``cloud'') outside the object. This cloud has, in general, a time-varying quadrupole moment and will slowly dissipate through emission of gravitational waves. On very long timescales, the end product is an object spinning so slowly that the instability is no longer active.

Because angular-momentum extraction occurs on a time scale $\tau=1/\omega_I$, the observation of an isolated compact object with spindown time scale $\tau_{\rm spindown}$ excludes superradiant instabilities for that system, at least on time scales $\tau<\tau_{\rm spindown}$. Therefore, compact objects for which a (possibly small) spindown rate can be measured accurately are ideal candidates to constrain the mechanism and, in turn, the dark-sector models discussed here. 

Unfortunately, measurements of the spin derivative of black holes are not available, so that constraints on superradiant instabilities using black-hole mass and spin measurements are only meaningful in a statistical sense~\cite{Brito:2014wla,Arvanitaki:2014wva,Arvanitaki:2016qwi}. On the other hand, both the spin and the spindown rate of pulsars are known with astonishing precision through pulsar timing (cf., e.g., Ref.~\cite{lorimer2005handbook}). For several sources, the rotational frequency is moderately high, $f_{\rm spin}=\Omega/(2\pi)\simeq (500-700)\,{\rm Hz}$, and the spindown time scale can be extremely long, $\tau_{\rm spindown} = \Omega/(\dot\Omega)\simeq 10^{10}\,{\rm yr}$. As an example, the ATNF Pulsar Catalogue~\cite{catalogue,Manchester:2004bp} contains $398$ ($40$) pulsars for which $\tau_{\rm spindown}>2\times 10^9\,{\rm yr}$ ($\tau_{\rm spindown}>2\times 10^{10}\,{\rm yr}$).

\begin{figure}[th]
  \includegraphics[width=0.48\textwidth]{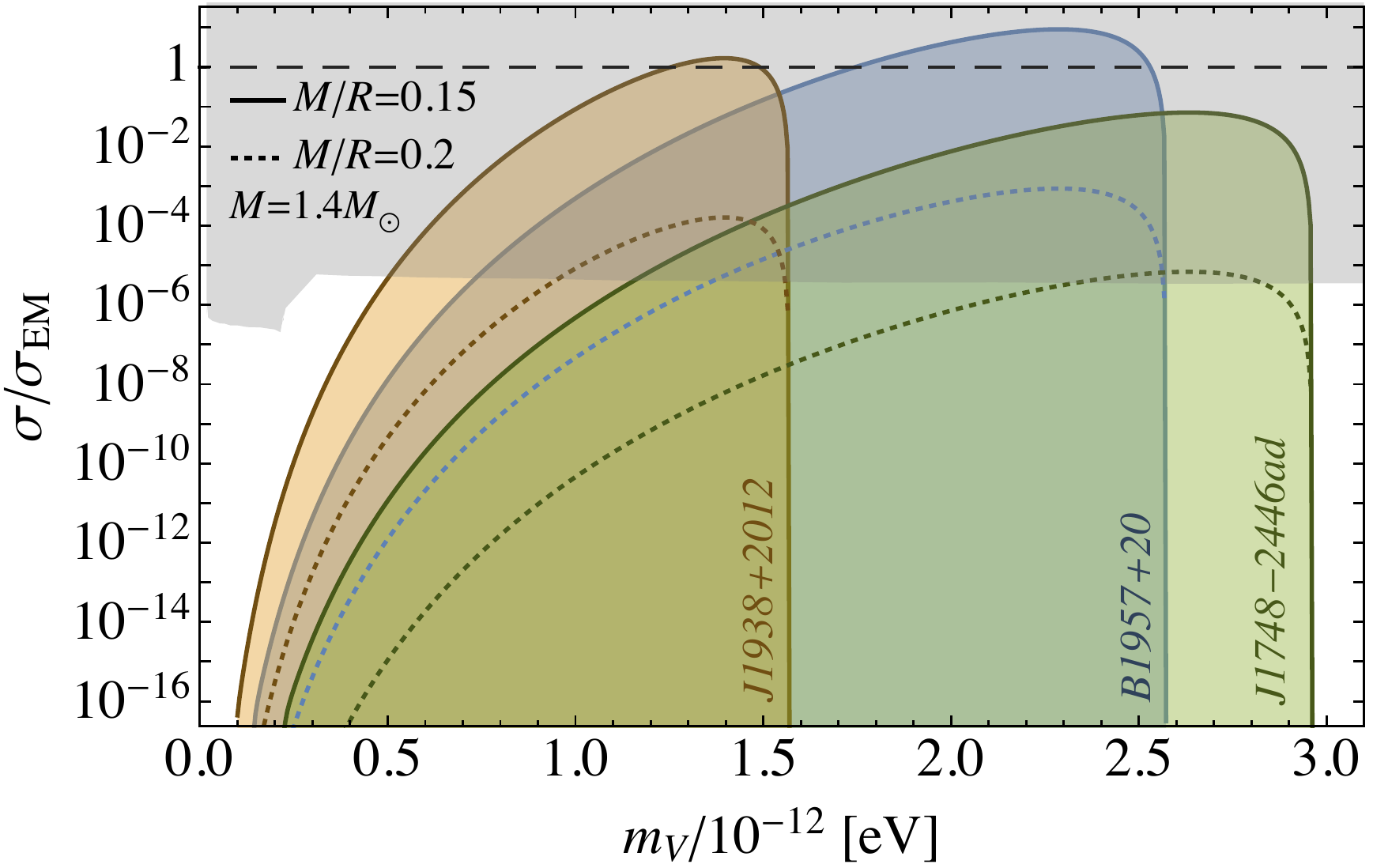}
\caption{Exclusion plots in the $\sigma/\sigma_{\rm{EM}}$ vs. $m_V$ plane obtained from the measurements of spin and spindown rate of pulsars J1938+2012 (orange)~\cite{Stovall:2016unz} and J1748-2446ad (green)~\cite{Hessels:2006ze}, and of the pulsar binary B1957+20 (blue)~\cite{Arzoumanian:1993qt}. In all cases we assumed $M= 1.4M_\odot$ and two values of the compactness, namely $M/R= 0.15$ (solid) and $M/R= 0.2$ (dotted). %
The shaded areas correspond to regions excluded by the superradiant instability because $\tau<\tau_{\rm spindown}$ for a given pulsar (i.e., the pulsar is observed to spin down at much longer rate than that predicted by the superradiant instability in that region of the parameter space). 
The horizontal dashed line corresponds to where $\sigma=\sigma_{\rm EM}$.
We only display the region where $\sigma\gg \Omega$. In the opposite limit, the instability time scale grows as $\tau\sim1/\sigma$ [cf.\ Eq.~\eqref{wIaxial}] and eventually $\tau>\tau_{\rm spindown}$ for sufficiently small $\sigma$, cf.\ discussion in the main text. 
The shaded gray region is excluded for $\sigma_\epsilon$ from distortions of the CMB blackbody from $\gamma\to X$ photon depletion~\cite{Mirizzi:2009iz}.
} \label{fig:pulsar}
%\end{center}
\end{figure}

In Fig.~\ref{fig:pulsar}, we show the excluded regions in the \emph{conductivity vs dark-photon mass plane} obtained by imposing $\tau<\tau_{\rm spindown}$ for three known sources, namely pulsars J1938+2012~\cite{Stovall:2016unz} and J1748-2446ad~\cite{Hessels:2006ze}, and pulsar binary B1957+20~\cite{Arzoumanian:1993qt}. The first one is representative of a pulsar with an exceptionally long spindown time scale ($\tau_{\rm spindown}\simeq1.1\times 10^{11}\,{\rm yr}$), but with a moderately large spin ($f_{\rm spin}\simeq 380\,{\rm Hz}$, which corresponds to $\Omega/\Omega_K\approx0.28$ assuming $M=1.4 M_\odot$ and $M/R=0.15$). 
The second one is the fastest pulsar known to date ($f_{\rm spin}\simeq 716\,{\rm Hz}$, corresponding to $\Omega/\Omega_K\approx0.53$ for $M=1.4 M_\odot$ and $M/R=0.15$), but only an upper bound on its spin derivative is available, from which we infer $\tau_{\rm spindown}>7.6\times 10^7\,{\rm yr}$.
The last one is representative of a pulsar with very large spin ($f_{\rm spin}\simeq 622\,{\rm Hz}$, which corresponds to $\Omega/\Omega_K\approx0.46$ again assuming $M=1.4 M_\odot$ and $M/R=0.15$), but moderately long spindown time scale ($\tau_{\rm spindown}\simeq3\times 10^{9}\,{\rm yr}$).
Furthermore, because our fits for $\alpha_1$ and $\alpha_2$ appearing in Eq.~\eqref{wIaxial} are independent of $\Omega$ only for $\Omega/\sigma\ll1$, in Fig.~\ref{fig:pulsar} we show only values of the conductivity which satisfy $\sigma\gg \Omega$. 

The exclusion plot shown in Fig.~\ref{fig:pulsar} is obtained as follows.
For a given measurement of the spin frequency of a pulsar, $f_{\rm spin}$, we can estimate $\Omega$ and compute the instability time scale as a function of $\sigma$ and $\mu_V$ through Eq.~\eqref{wIaxial}. Furthermore, the measurement of a spindown timescale for a pulsar, $\tau_{\rm spindown}$, implies that a faster spindown rate caused by the superradiant instability would be incompatible with observations.
Thus, imposing $\tau<\tau_{\rm spindown}$ yields an excluded region in the $\sigma$-$m_V$ plane. Fastly spinning pulsars constrain the rightmost part of the $\sigma$-$m_V$ diagram because the instability requires $\mu_V\sim\omega_R <m\Omega$. On the other hand, pulsars with longer spindown time scale correspond to higher threshold lines in the leftmost part of the $\sigma$-$m_V$ diagram.

%%%%%%%%%%%%%%%%%%%%%%%%%%%%%%%%%%%%%%%%%%%%%%%%%%%%%%%%%%%%%%%%%%%%%%%%%%%%
\subsection{Superradiantly-induced maximum spin frequency for pulsars}
%%%%%%%%%%%%%%%%%%%%%%%%%%%%%%%%%%%%%%%%%%%%%%%%%%%%%%%%%%%%%%%%%%%%%%%%%%%%

Accreting neutron stars in the weakly magnetic Low-Mass X-Ray Binaries (LMXBs) are expected to be spun up near the mass-shedding frequency in a spinup time scale
%%%
\begin{equation}
 \tau_{\rm spinup}\sim 10^8\left(\frac{10^9 M_\odot\,{\rm yr}^{-1}}{\dot M}\right)\, {\rm yr} \,,\label{spinup}
\end{equation}
where $\dot M$ is the mass-accretion rate. Since the above time scale is much less than the age of a typical LMXB, many accreting neutron stars in weakly magnetic LMXBs should be
observed rotating near the mass-shedding frequency, $\Omega_K/(2\pi)\gtrsim 1\,{\rm kHz}$.
The lack of observed systems with $f_{\rm spin}\gtrsim700\,{\rm Hz}$ has motivated various limiting mechanisms for the maximum spin of a pulsar, many of them involving gravitational-wave dissipation --~either through an accretion-induced mass quadrupole on the crust~\cite{Bildsten:1998ey}, a large toroidal magnetic
field~\cite{Cutler:2002nw}, or through the excitation of the unstable r-modes~\cite{Andersson:1998ze,Andersson:1997xt}~-- and more recently advocating the disk/magnetosphere interaction as leading spindown mechanism~\cite{Patruno:2011sj}.

One might wonder whether --~besides placing direct constraints on models of dark photons~-- the superradiant instability of neutron stars could also provide an alternative (albeit exotic) explanation for the spin limit of observed pulsars.
For fixed values of $\sigma$ and $\mu_V$, our model predicts that an accreting pulsar in a LMXB (for which the superradiant instability is initially effective) would reach a critical angular velocity such that $\tau(\Omega)=\tau_{\rm spinup}$ in a small fraction of its age. 
However, because $\tau_{\rm spinup}\ll\tau_{\rm spindown}$ for the observed pulsars discussed in the previous sections, the threshold line $\tau=\tau_{\rm spinup}$ is already excluded by pulsar timing. 
In other words, only models that are already excluded by Fig.~\ref{fig:pulsar} would produce a superradiant instability strong enough to overcome accretion at a time scale given by Eq.~\eqref{spinup}.
%%%%%%%%%%%%%%%%%%%%%%%%%%%%%%%%%%%%%%%%%%%%%%%%%%%%%%%%%%%%
\section{Discussion and future work}
%%%%%%%%%%%%%%%%%%%%%%%%%%%%%%%%%%%%%%%%%%%%%%%%%%%%%%%%%%%%
The scattering of light by rotating, conducting spheres is a classical problem in electromagnetism, and can lead to superradiant effects. 
Yet, to the best of our knowledge, a thorough understanding of this problem has not been framed within the context of superradiance.
Superradiance in stars may have interesting and important applications in astrophysics and particle physics:
stars are made of materials with small but nonvanishing resistivity in the standard Maxwell sector, leading to the amplification of
low-energy pulses. In the context of ultralight dark photon models, any nonzero conductivity of stars in the dark sector
will lead to superradiant instabilities that drive the star to lower rotation rates. In other words, the superradiant mechanism leads to potentially observable consequences,
which can be used to constrain the dark sector. 

We have shown that a direct signature of superradiant instabilities in stars is the spindown of pulsars in the presence of ultralight dark photons. As we discussed, existing measurements of the spindown rate of pulsars already place some stringent constraint on models of dark photons and of the hidden $U(1)'$ sectors.
Although superradiance is typically weaker for stars than for black holes, the spindown rate of pulsars is measured with great precision and it is typically very low (i.e., $\tau_{\rm spindown}$ is very long), leading to \emph{direct} constraints which are much more robust than those coming from mass-spin distributions in the so-called black-hole Regge plane.
For example, our preliminary analysis suggests that ordinary models ($\sigma\lesssim \sigma_{\rm EM}$) of dark photons with mass $m_V\sim 10^{-12}\,{\rm eV}$ are excluded by pulsar-timing observations.  

There are many interesting follow-up questions to the effect of superradiance in stars. One of them concerns the polar sector of vector perturbations. Previous studies of black hole superradiance
show that the vector sector triggers instabilities with much shorter time scales~\cite{Pani:2012vp,Pani:2012bp,Witek:2012tr}. If such a result generalizes to conducting stars, the constraints on dark photons will certainly improve. 
We hope that the promising results of our exploratory study shown in Fig.~\ref{fig:pulsar} will stimulate further investigation on this problem, including a complete analysis of the constraints that can be placed on dark-photon models with pulsar timing.
From a theoretical perspective, another interesting open issue concerns the functional dependence of the amplification factor on the frequency. Previously, effective field theory approaches have investigated the frequency dependence in the context of black holes~\cite{Endlich:2016jgc}. It would be interesting to extend such an approach to stars.
Furthermore, in this work we modelled the conductivity with a simple Drude model, in which electrons only scatter with protons. This gives us an order of magnitude
of the constraints that one can impose via superradiance, and motivates a more complete calculation (e.g.~\cite{1970ApJ...159..641C,1976ApJ...206..218F}). 

In the scenario in which the dark photon couples to Maxwell vectors, superradiance could work in more intricate ways: on the one hand both vectors are superradiantly amplified by the star's material, potentially leading to a stronger effect. On the other hand, Maxwell fields are massless
and could easily escape, not being subject to the confinement necessary to create the instability. How exactly the mechanism proceeds depends on this interplay and depends on more detailed calculations.
Furthermore, it would be interesting to explore the coupling to plasma. Equation~\eqref{plasma_freq} shows that even ordinary photons would acquire an effective mass $\hbar\omega_p \sim 10^{-12}\,{\rm eV}$ when propagating in a plasma with electron number density $n_{\rm plasma}\sim 10^{-2}\,{\rm cm}^{-3}$. This might give interesting superradiant effects for ordinary photons~\cite{Pani:2013hpa,Conlon:2017hhi} or also alter the instability for dark photons if the latter are coupled to plasma sufficiently strongly.

Our analysis also shows that it is, in principle, possible to generalize a number of results in the literature concerning black hole superradiance~\cite{Brito:2015oca}.
For example, for complex, massive vector fields there should exist new stationary solutions describing a star surrounded by a Proca condensate.
This would be a natural generalization of the hairy black hole solutions found recently~\cite{Herdeiro:2014goa,Herdeiro:2015waa,Herdeiro:2016tmi}.
Likewise, imprints of superradiance in the luminosity of pulsars or black hole binaries~\cite{Rosa:2015hoa,Rosa:2016bli} should also be present when the companion is a star, instead of a black hole.
Finally, the development of the instability will certainly lead to nontrivial gravitational-wave emission. In the black hole case, the emitted signal can be used to impose interesting constraints on the models~\cite{Arvanitaki:2014wva,Brito:2014wla,Yoshino:2014wwa,Arvanitaki:2016fyj,Brito:2015oca}. On the other hand, stars have typically lower masses than black holes, and it remains to be understood if gravitational-wave emission is relevant in this case.

%%%%%%%%%%%%%%%%%%%%%%%%%%%%%%%%%%%%%%%%%%%%%%%%%%%%%%%%%%%%%%%%%%%%%%%%%%%%%%
%\noindent{\bf{\em Acknowledgments.}}
%%%%%%%%%%%%%%%%%%%%%%%%%%%%%%%%%%%%%%%%%%%%%%%%%%%%%%%%%%%%%%%%%%%%%%%%%%%%%%
\begin{acknowledgments}
We are indebted to Leonardo Gualtieri for suggesting a possible connection to the spin limit of pulsars, and to Masha Baryakhtar, Sam Dolan, Mauricio Richartz, and Jo\~ao Rosa for providing useful discussions and comments on the draft.
V.C. acknowledges financial support provided under the European Union's H2020 ERC Consolidator Grant ``Matter and strong-field gravity: New frontiers in Einstein's theory'' grant agreement no. MaGRaTh--646597. Research at Perimeter Institute is supported by the Government of Canada through Industry Canada and by the Province of Ontario through the Ministry of Economic Development $\&$
Innovation.
This project has received funding from the European Union's Horizon 2020 research and innovation programme under the Marie Sklodowska-Curie grant agreement No 690904 and from FCT-Portugal through the projects IF/00293/2013.
The authors would like to acknowledge networking support by the COST Action CA16104.
The authors thankfully acknowledge the computer resources, technical expertise and assistance provided by S\'ergio Almeida at CENTRA/IST. Computations were performed at the cluster ``Baltasar-Sete-S\'ois'', and supported by the MaGRaTh--646597 ERC Consolidator Grant.
\end{acknowledgments}
%%%%%%%%%%%%%%%%%%%%%%%%%%%%%%%%%%%%%%%%%%%%%%%%%%%%%%%%%%%%%%%%%%%%%%%%%%%%%%
% \clearpage
% \newpage

\appendix

%%%%%%%%%%%%%%%%%%%%%%%%%%%%%%%%%%%%%%%%%%%%%%%%%%%%%%%%%%
\section{Thin-shell model and membrane paradigm}\label{app:membrane}
%%%%%%%%%%%%%%%%%%%%%%%%%%%%%%%%%%%%%%%%%%%%%%%%%%%%%%%%%%
In the membrane paradigm~\cite{Thorne:1986iy}, a black hole can be interpreted as a one-way membrane endowed with various properties. In particular, the surface resistivity reads $R_H=4\pi\simeq 377\,{\rm Ohm}$, so that the \emph{surface conductivity} is $\hat\sigma = 1/(4\pi)$.

Within our framework, a similar model can be investigated by considering a conducting thin shell in vacuum, so that the (volume) conductivity reads $\sigma(r) =\hat\sigma\delta(R-r)$. For simplicity, we consider the Newtonian limit, in which $F=B=1$, and restrict ourselves to small frequencies, so that $\omega\zeta$ in Eq.~\eqref{axial_final} is negligible. In these approximations, axial perturbations reduce to Bessel's equation
%%%
\begin{equation}
 \frac{d^2a}{dr^2}+\left(\omega^2-\frac{l(l+1)}{r^2}\right)a=0\,,
\end{equation}
%%%
both in the interior and in the exterior. The delta function in $\sigma(r)$ enters only in the junction conditions, which imply
%%%
\begin{equation}
 [[da/dr]]=-4\pi i \hat\sigma (\omega-m\Omega) a(R)\,,
\end{equation}
%%%
where $[[...]]$ is the jump across the shell and, without loss of generality, we assumed $[[a]]=0$. We impose the junction condition above on the solutions of the Bessel's equation with correct boundary conditions as discussed in the main text. For $l=1$, we obtain
%%%
\begin{equation}
 Z=-\frac{16\pi}{9} \hat \sigma  (R\omega)^3 R(\omega-m\Omega)\,,
\end{equation}
%%%
where $m=1,0,-1$. In the nonrotating case, this result is valid also beyond the small-frequency regime and, interestingly, it agrees exactly with that obtained in black hole perturbation theory [cf.\ Ref.~\cite{Brito:2015oca}, Eq.~(3.103)] upon identification of $\hat\sigma=1/(4\pi)$ and $R=2M$. Thus, a by-product of our analysis is the proof that the black hole membrane paradigm works also for linear electromagnetic perturbations.

The shell toy-model is also useful to understand the results for the instability. Instead of a massive field, we consider a spinning shell of radius $R$ surrounded by a nonspinning perfect conductor of radius $R_2$. The characteristic modes of the system can be found by imposing the above junction condition and $a(r=R_2)=0$. For large values of $R_2/R$, the $l=1$ fundamental mode reads
\beq
\omega &=& \frac{\gamma_0}{R_2} -4 \pi i  \gamma_0 \left(\gamma_0^2+1\right)\frac{R^4}{R_2^5}\hat\sigma  \Upsilon\,,
\eeq
where $\gamma_0$ satisfies $\tan\gamma_0=\gamma_0$ and
%%%
\begin{equation}
 \Upsilon =  \frac{ \gamma_0 \left(9-16 \pi ^2 \hat\sigma ^2 \Omega^2 R^2\right)- \Omega R_2 \left(9+16 \pi ^2 R^2 \hat\sigma ^2 \Omega^2\right)}{ \left(9+16 \pi ^2  \hat\sigma ^2 \Omega ^2 R^2\right)^2}\,.
\end{equation}
%%%
%%%
Note that $\Upsilon\to (\gamma_0-\Omega R_2)/9$ when $\hat\sigma \Omega R\ll1$ (thus recovering the superradiant condition, $\omega_R<\Omega$), whereas $\Upsilon\sim - 1/(\hat\sigma^2 \Omega^2 R^2) $ when $\hat\sigma \Omega R\gg1$. 
At fixed rotation rate, the peak of the instability occurs at $\hat\sigma\sim 3/(4\pi \Omega R)$.
Therefore --~at least qualitatively~-- this simple model shows the same features that we observe numerically, in particular the fact that the instability decreases as $\hat\sigma\to0$ and at very large $\hat\sigma$, and it also informs us on the $\Omega$ dependence. Finally, if we substitute $R_2\to 1/\mu_V^2$ as discussed in Refs.~\cite{Brito:2014nja,Brito:2015oca}, we recover the mass dependence presented in the main text for $\mu_V\ll \Omega$.

%%%%%%%%%%%%%%%%%%%%%%%%%%%%%%%%%%%%%%%%%%%%%%%%%%%%%%%%%%
\section{Vector perturbations of a slowly-spinning compact object}\label{app:slowrot}
%%%%%%%%%%%%%%%%%%%%%%%%%%%%%%%%%%%%%%%%%%%%%%%%%%%%%%%%%%
In this appendix we follow the framework developed in Refs.~\cite{Kojima:1992ie,Pani:2012vp,Pani:2012bp,Pani:2015hfa,Pani:2015nua} (cf.\ Ref.~\cite{Pani:2013pma} for a review) to derive the Proca perturbations of a slowly-rotating, conducting star.
The Proca equation~\eqref{proca}, linearized in the perturbations~\eqref{expansion} on the background~\eqref{metric} can be written in the following form\footnote{We will append the relevant multipolar index ${l}$ to any perturbation variable but we will omit the index $m$, because in an axisymmetric background it is possible to decouple the perturbation equations so that all quantities have the same value of $m$.
}:
%%%
\begin{eqnarray}
 \delta\Pi_{I}&\equiv&  \left(A^{(I)}_{{l}}+{\tilde A}^{(I)}_{{l}}\cos\th\right)Y^{l} +B^{(I)}_{{l}}\sin\th\pa_{\th}Y^{l}=0\,,\nn\\\label{eq1i}\\ 
%%%
 \delta\Pi_{\vartheta}&\equiv&   \alpha_{{l}}
\pa_{\th}Y^{l}-{i} m\beta_{l}\frac{Y^{l}}{\sin\th}+\eta_{l}\sin\th Y^{l}=0\,, \label{eq2}\\
 \frac{\delta\Pi_{\varphi}}{\sin\th}&\equiv& \beta_{{l}}\pa_{\th}Y^{l}+{i} m\alpha_{l}\frac{Y^{l}}{\sin\th} +\zeta_{l}\sin\th Y^{l}=0\,, \label{eq3}
\end{eqnarray}
%%%%
where a sum over $({l},m)$ is implicit and $I$ denotes either the $t$
component or the $r$ component.  The various radial coefficients in Eqs.~\eqref{eq1i}--\eqref{eq3} are given in a supplemental {\scshape Mathematica}\textsuperscript{\textregistered} notebook~\cite{webpages}.
Each of these coefficients is a linear
combination of perturbation functions with either polar or axial
parity. Therefore we can divide them into two sets:
\begin{eqnarray}
 &&\text{Polar:}\qquad A^{(I)}_{{l}}\,,\quad \alpha_{{l}}\,,\quad \zeta_{{l}}\,,\nn\\
%%%
 &&\text{Axial:}\qquad \tilde A^{(I)}_{{l}}\,,\quad B^{(I)}_{{l}}\,,\quad 
\beta_{{l}}\,,\quad \eta_{{l}}\,,\nn
\end{eqnarray}
%%%
where $I=t,r$. 

%%%%%%%%%%%%%%%%%%%%%%%%%
\subsection{Separation of the angular dependence}
%%%%%%%%%%%%%%%%%%%%%%%%%

In order to separate the angular variables in
Eqs.~\eqref{eq1i}--\eqref{eq3} we compute the following integrals:
\begin{subequations}
\begin{align}
&\int\delta\Pi_I Y^{*\,{{l}}}d\Omega\,,\quad (I=t,r)\,;\\
&\int\delta\Pi_a \mathbf{Y}_b^{*\,{{l}}}\gamma^{ab}d\Omega\,,\quad (a\,,b=\vartheta,\varphi)\,;\\
&\int\delta\Pi_a \mathbf{S}_b^{*\,{{l}}}\gamma^{ab}d\Omega\,,\quad (a\,,b=\vartheta,\varphi)\,.
\end{align}
\end{subequations}
%%%
where we set $x^{\mu}=(t,r,x^b)$ with $x^b=(\vartheta,\varphi)$, the two-sphere $\gamma_{ab}={\rm
  diag}(1,\sin^2\vartheta)$, and
\begin{eqnarray}
\mathbf{Y}_b^{{l}}&=&\left(\partial_\vartheta Y^{{l}},\partial_\varphi Y^{{l}}\right)\,,\nonumber\\
\mathbf{S}_b^{{l}}&=&\left(\frac{1}{\sin\vartheta}\partial_\varphi Y^{{l}},
-\sin\vartheta\partial_\vartheta Y^{{l}}\right)\,.
\end{eqnarray}
%%%
%
We also make use of the orthogonality properties of scalar and vector harmonics, namely
\begin{align}
&\int Y^{{l}} Y^{*\,{l}'}d\Omega=\delta^{{l}{l}'}\,,\\
&\int \mathbf{Y}_b^{{l}} \mathbf{Y}_b^{*\,{l}'}\gamma^{ab} d\Omega=\int \mathbf{S}_b^{{l}} 
\mathbf{S}_b^{*\,{l}'}\gamma^{ab} d\Omega={{l}({l}+1)}\delta^{{l}{l}'}\,,\nn\\
&\int \mathbf{Y}_b^{{l}} \mathbf{S}_b^{*\,{l}'}\gamma^{ab} d\Omega=0\,,
\end{align}
as well as of the identities
\begin{eqnarray}
\cos\th Y^{{l}}&=&\cQ_{{l}+1}Y^{{l}+1}+\cQ_{{l}}Y^{{l}-1}\,,\label{ident1}\\
\sin\th \partial_\vartheta Y^{{l}}&=&
{\cal Q}_{{l}+1}{l} Y^{{l}+1}-{\cal Q}_{{l}}({l}+1)Y^{{l}-1}\,,\label{ident2}
\end{eqnarray}
% %%
with ${\cal Q}_{l}=\sqrt{\frac{{l}^2-m^2}{4{l}^2-1}}$.
By using the above relation, we obtain the following \emph{radial} equations:
\begin{align}
&A^{(I)}_{{{l}}}+\cQ_{{{l}}}\left[{\tilde A}^{(I)}_{{l}-1}+({l}-1){B}^{(I)}_{{l}-1}\right]\nn\\
&+\cQ_{{l}+1}\left[{\tilde A}^{(I)}_{{l}+1}
-({l}+2){ B}^{(I)}_{{l}+1}\right]=0\,,\label{eqset1}\\
&{{l}({l}+1)}\alpha_{{l}}-{i} m\zeta_{{l}}\nn\\
&-{\cal Q}_{{l}}({l}+1)\eta_{{l}-1}+{\cal Q}_{{l}+1}{l} \eta_{{l}+1}=0\,,\label{eqset2}\\
&{{l}({l}+1)}\beta_{{l}}+{i} m\eta_{{l}}\nn\\
&-{\cal Q}_{{l}}({l}+1)\zeta_{{l}-1}+{\cal Q}_{{l}+1}{l} \zeta_{{l}+1}=0\,.\label{eqset3}
\end{align}
%%%%%
Note that Eqs.~\eqref{eqset1}--\eqref{eqset3} can be written in the schematic form
\begin{eqnarray}
0&=&{\cal A}_{{l}}+\epsilon_a m \bar{\cal A}_{{{l}}}+\epsilon_a ({\cal Q}_{{{l}}}\tilde{\cal P}_{{l}-1}+{\cal Q}_{{l}+1}\tilde{\cal P}_{{l}+1})\,,\label{eq_axial}\\
%%%%%
0&=&{\cal P}_{{l}}+\epsilon_a m \bar{\cal P}_{{{l}}}+\epsilon_a ({\cal Q}_{{{l}}}\tilde{\cal A}_{{l}-1}+{\cal Q}_{{l}+1}\tilde{\cal A}_{{l}+1})\,,\label{eq_polar}
\end{eqnarray}
%%%%%
where $\epsilon_a$ is a bookkeeping parameter for the expansion in the angular momentum,  ${\cal A}_{{l}}$, $\bar {\cal A}_{{l}}$ are linear combinations of the axial perturbations with multipolar index ${l}$; similarly, ${\cal P}_{{l}}$, $\bar {\cal  P}_{{l}}$ are linear combinations of the polar perturbations with index ${l}$.

%%%%%%%%%%%%%%%%%%%%%%%%%
\subsection{Axial-led and polar-led perturbations}
%%%%%%%%%%%%%%%%%%%%%%%%%
We expand the axial and polar perturbation functions (schematically denoted as $a_{{l}}$ and $p_{{l}}$, respectively) that appear in Eqs.~\eqref{eq_axial} and \eqref{eq_polar} as
%%%
\begin{eqnarray}
a_{{l}}&=&a^{(0)}_{{l}}+\epsilon_a\,a^{(1)}_{{l}}+{\cal O}(\epsilon_a^2)\nn\\
p_{{l}}&=&p^{(0)}_{{l}}+\epsilon_a\,p^{(1)}_{{l}}+{\cal O}(\epsilon_a^2)\,.
\end{eqnarray}
%%%%
Since in the nonrotating limit axial and polar perturbations are
decoupled, a possible consistent set of solutions of the system
\eqref{eq_axial}--\eqref{eq_polar} has $p^{(0)}_{{L}\pm1 }\equiv0$, where ${l}=L$ is a specific value of the harmonic index. This ansatz leads to the so-called ``axial-led'' subset of Eqs.~\eqref{eq_axial}--\eqref{eq_polar}:
\begin{equation}\label{axial_led}
 \left\{ \begin{array}{l}
          {\cal A}_{{L}}+\epsilon_a m \bar{\cal A}_{{{L}}}=0\\
{\cal P}_{{L}+1}+
\epsilon_a {\cal Q}_{{{L}+1}}\tilde{\cal A}_{{L}}=0\\
{\cal P}_{{L}-1}+
\epsilon_a {\cal Q}_{{{L}}}\tilde{\cal A}_{{L}}=0
         \end{array}\right.\,, 
\end{equation}
%%%%%
where the first equation is solved to first order in the spin, whereas the second and the third equations do not contain zeroth-order quantities in the spin, i.e. $p_{L\pm1}={\cal O}(\epsilon_a)$. The truncation above is consistent because in the axial equations for ${l}=L$ the polar source terms with ${l}=L\pm1$ appear multiplied by a factor $\epsilon_a$, so they would enter at second order in the rotation.

Similarly, another consistent set of solutions of the same system has $a^{(0)}_{{L}\pm1}\equiv0$. The corresponding ``polar-led'' system reads 
\begin{equation}\label{polar_led}
 \left\{ \begin{array}{l}
          {\cal P}_{{L}}+\epsilon_a m \bar{\cal P}_{{{L}}}=0\\
{\cal A}_{{L}+1}+\epsilon_a {\cal Q}_{{{L}+1}}\tilde{\cal P}_{{L}}=0\\
{\cal A}_{{L}-1}+\epsilon_a {\cal Q}_{{{L}}}\tilde{\cal P}_{{L}}=0
% \\
         \end{array}\right.\,.
\end{equation}
%%%%%%
Interestingly, within this perturbative scheme a notion of ``conserved quantum number'' ${L}$ is still meaningful: even though, for any given ${L}$, rotation couples terms with opposite parity and different multipolar index, the subsystems~\eqref{axial_led} and \eqref{polar_led} are closed, i.e. they contain a \emph{finite} number of equations which describe the dynamics to first order in the spin. 

Finally, note that the first set of equations in the axial-led system~\eqref{axial_led} and in the polar-led system~\eqref{polar_led} do not involve couplings between axial and polar modes. Once the first set of equations in the system~\eqref{axial_led} [or in the system~\eqref{polar_led}] is solved, the remaining two equations can be solved separately. Therefore, if one is interested in the linear spin corrections to axial or polar perturbations with a given harmonic index $L$, one can solve only the first set of equations in the system~\eqref{axial_led} or~\eqref{polar_led}, respectively.

%%%%%%%%%%%%%%%%%%%%%%%%%
\subsubsection{Final equations for the axial-led system}
%%%%%%%%%%%%%%%%%%%%%%%%%

By using the coefficients given in the Supplemental Material, it is easy to show that the first equation of the system~\eqref{axial_led} reduces to Eq.~\eqref{axial_final} in the main text.

%%%%%%%%%%%%%%%%%%%%%%%%%
\subsubsection{Final equations for the polar-led system}
%%%%%%%%%%%%%%%%%%%%%%%%%
The polar-led system is more involved. In general, one of the polar equations fixes the proper charge density $\hat\rho_{\rm EM}$ in terms of the other perturbation functions, even when $\sigma={\rm const}$.
In the Proca case, by using the coefficients in the Supplemental Material, the system can be reduced to three differential equations that can be schematically written as
%%%%
\begin{eqnarray}
 u_1''&=& f_1(u_1,u_1',u_2)\,,\nn \\
 u_2'&=& f_2(u_1,u_1',u_2,u_3)\,,\nn \\
 u_3'&=& f_3(u_1',u_2,u_3)\,,\nn 
\end{eqnarray}
%%%%
where $u_1\equiv f^{lm}$ , $u_2\equiv h^{lm}$ and $u_3=k^{lm}$.
Note that the first equation above does not contain $u_3$. Therefore, it is possible to write a system of two second-order, radial equations for $u_1$ and $u_2$ simply by solving the second equation above for $u_3$, differentiate it with respect to $r$, and then using the third equation above to eliminate $u_3$. The final result is not shown explicitly and a detailed investigation is left for future work.
Note that in the Maxwell case ($\mu_V=0$) the usual gauge freedom can be used to eliminate one spurious degree of freedom. Consequently, the Maxwell polar sector propagates only one degree of freedom, described by a second-order field equation.

\bibliographystyle{apsrev4}
\bibliography{Ref}

\end{document}